\documentclass[twocolumn,superscriptaddress,showpacs,pra,floatfix]{revtex4}
\usepackage{graphicx}
\usepackage{amsmath}
\usepackage{amssymb}
\usepackage{amsfonts}
\usepackage{subeqnarray}
\usepackage{subfigure}
\usepackage{bm}

\newcommand{\mc}{\mathcal} 
\newcommand{\ket}[1]{ \left| #1 \right\rangle }

\newcommand{\Imag}[1]{\textrm{Im}\left[#1\right]}
\begin{document}

\title{Photon Wave-packet Manipulation via Dynamic Electromagnetically Induced Transparency in Multilayer Structures}

\author{Francesco Bariani}
\email{bariani@science.unitn.it}
\affiliation{CNR-INFM BEC Center and Dipartimento di Fisica, Universit`a di Trento, I-38050 Povo, Italy} 

\author{Iacopo Carusotto}
\affiliation{CNR-INFM BEC Center and Dipartimento di Fisica, Universit`a di Trento, I-38050 Povo, Italy} 

\begin{abstract}
We present a theoretical study of the dynamics of a light pulse propagating through a multilayer system consisting of alternating blocks of electromagnetically induced transparency (EIT) media and vacuum. We study the effect of a dynamical modulation of the EIT control field on the shape of the wave packet. Interesting effects due to the group velocity mismatch at the interfaces are found. Modulation schemes that can be realized in ultracold atomic samples with standard experimental techniques are proposed and discussed. Calculations are performed using a modified slowly varying envelope approximation of the Maxwell-Bloch equations and are compared to an effective description based on a continuity equation for the polariton flow. 
\end{abstract}


\date{\today}
\maketitle
\section{Introduction}
% Coherence in radiation-matter interaction
The control of light-pulse propagation in matter is a key element of optical devices for fundamental science as well as for technological applications. In many cases, this is made difficult by the presence of competing effects like dispersion and absorption. Furthermore, the available time for manipulation is often limited by the very high propagation speed of light in conventional materials.

New perspectives in light propagation are opened by the observation of long-living coherence effects in optical media. The Coherent Population Trapping (CPT) \cite{CPT_review} and Electromagnetically Induced Transparency (EIT) \cite{EIT_review} effects have been shown to produce strong modification to the properties of optical media.
By dressing the matter excitations with coherent external fields, a resonant probe laser pulse can be made to propagate across an otherwise strongly absorbing medium at an ultraslow group velocity and without being distorted. The incoming light is coupled to a dark-state polariton which shows vanishing absorption and dispersion \cite{EIT_dark_polariton} and whose group velocity can be controlled via the intensity of the control field \cite{Slow_light_review}. This feature has been demonstrated as a tunable delay lines for propagating pulses.

% Dynamical modulation of EIT
The dynamical modulation in time of the control field while the pulse is propagating opens up even richer possibilities for light manipulation in the spirit of the so-called Dynamic Photonic Structures \cite{DPS,EIT_spatial}. 
For example, by completely switching off the control field, the probe light can be halted and stored as an atomic (spinlike) excitation, and later retrieved after a macroscopic time: such light-storage techniques \cite{EIT_dark_polariton,optimal_light_storage} are considered crucial tools for all-optical information technologies.
A periodic dynamical modulation of a spatially homogeneous control field can lead to intriguing phenomena such as frequency triggering in time of the EIT band \cite{EIT_phase}. By adiabatically raising the resonance frequency, it is also possible to shift the energy of a propagating polariton \cite{bariani}. A nonadiabatic variation of the control field has been proposed as a tool to compensate the pulse broadening at the exit of a delay line \cite{instant_modulation} or after retrieval of a previously stored light wave packet \cite{retrieved_shaping}. Extremely fast modulations of the control field have been anticipated to produce a substantial dynamical Casimir emission \cite{carusotto_casymir}.

% Spatial modulation of EIT
Spatially modulated EIT media have been proposed and used for light-stopping applications, via the creation of tunable stop bands within the EIT window \cite{EIT_spatial,EIT_spatial2}. Mutual interaction of several moving spin coherence gratings has been demonstrated as an efficient way to stop two-color light and to perform wavelength conversion \cite{two_color_EIT} .

% Temporal shaping
In the present article we investigate the effect on ultraslow light propagation of a combined spatial and temporal modulation of the EIT medium.
A pioneering study in this direction was reported in \cite{EIT_shaping, CPT_shaping} where the simultaneous propagation of both control and probe pulses is investigated: a ramp of the control field in an otherwise homogeneous medium induces different propagation velocities in the different parts of the probe pulse. This is predicted to result in a controllable reshaping of the probe profile.

% Experiments in cold gases
Dilute ultracold gases are among the most promising media for EIT applications. Both slow light and light storage have been experimentally realized in this systems \cite{EIT_experiment}. As compared to condensed matter systems, atomic gases have the advantage of showing extremely narrow linewidths, which can be exploited to obtain ultraslow group velocities: Doppler and collisional dephasing processes which destroy the coherence are in fact strongly reduced in cold samples. Furthermore, the achievement of Mott-insulator (MI) phase \cite{MI} suggests the possibility of obtaining even longer coherence times thanks to the gap in the many-body excitation spectrum: the first experimental realization of EIT in a MI has been recently reported for light-storage purposes \cite{EIT_bloch}.

% Small samples: take advantages
% Effect of interfaces: peculiar spatial modulation of EIT medium
Most of the theoretical works on slow light propagation in atomic samples were focused on the simplest case of a homogeneous atomic medium with some boundary condition. Unfortunately, the typical size of atomic samples is often small as compared to the duration and waist of the probe pulse, which calls for a more complete theoretical description. In this article we present a model that is able to include the spatial inhomogeneity of a system and therefore to describe the propagation dynamics at the interface between vacuum and the EIT medium. 

In particular, we show how one can take advantage of the interfaces to manipulate the wave-packet shape by means of a dynamical modulation of the control field intensity. The multilayer structure offers in fact the possibility of spatially engineering the radiation-matter interaction. While the usual schemes based on a single control field pulse in a homogeneous system \cite{EIT_shaping} limit the manipulation to a single interface, our proposal can explore a much wider range of configurations: increasing the number of interfaces allows a variety of modulation protocols to be developed, leading, for example, to the switching of a single pulse into a train of separated signals. Furthermore, the reduced optical depth of each layer may allow for a more efficient modulation \cite{optimal_light_storage}. The short pulses that can be tailored with our proposed technique may be of interest in view of creating polariton states with a Dirac-like dispersion \cite{EIT_Dirac}. Even though our theoretical description is based on semiclassical wave equations, our conclusions directly apply to the quantum processing of single photon wavefunctions and therefore can have an importance for optical quantum computing applications.

% Outline
The article is organized as follows. In Sec. II we review the Maxwell-Bloch formalism for describing light propagation in an atomic medium and we introduce the modified Slowly Varying Envelope Approximation to include the spatial inhomogeneities. The standard polariton picture for slow light propagation in homogeneous EIT media is reviewed in Sec. III. These concepts are then used in Sec. IV to investigate the simplest modulation schemes based on the idea of EIT chain. In Sec. V, we derive an effective equation for the pulse propagation. This simplified formalism is then used in Sec. VI to propose and characterize some specific protocols to be implemented in cold atom systems. In Sec. VII, we draw conclusions and we sketch the perspectives of the work.

\section{The formalism}

This article is devoted to the theoretical analysis of the propagation of a light pulse in a time-dependent and spatially inhomogeneous medium with sharp interfaces. Furthermore, the pulse carrier is fixed in a region where the optical response of the medium is strongly dispersive. 
This requires that the formalism explicitly includes both the atomic degrees of freedom in the presence of a time-dependent dressing field, and the spatiotemporal dynamics of the propagating pulse. 
In this section we introduce a modified version of the well-known Slowly Varying Envelope Approximation which is able to address all these features in a single unified treatment.

\begin{figure*}[htbp]
\begin{center}
\includegraphics[width = 0.9\textwidth,clip]{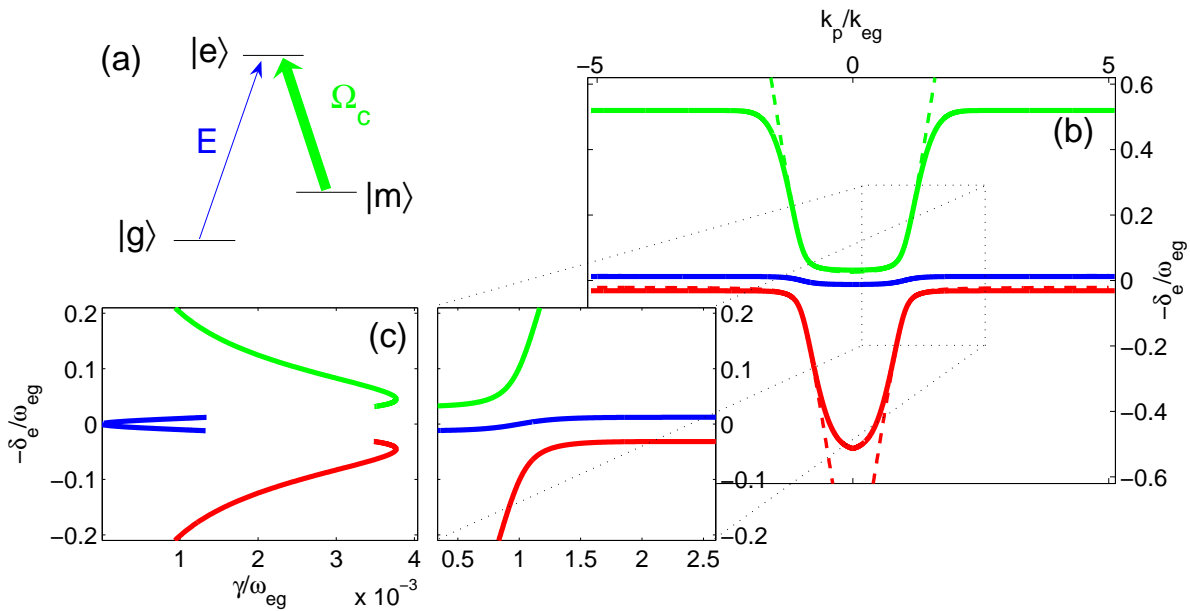}
\caption{(a) Scheme of the three-level $\Lambda$ configuration of atomic levels and laser beams. (b) and (c) Frequency dispersion and decay rate of the three polariton branches (green, upper; blue, dark; red, lower). Parameters of the system: $\sqrt{D} = 0.1$, $\Omega_c/\omega_{eg} = 0.04$, $\gamma_e/\omega_{eg} = 0.01$, resonant dressing $\omega_c = \omega_e-\omega_m$. The solid lines are the MSVEA predictions using $\omega_p = \omega_{eg}$ and and assuming an Erf shape for the $f(k)$ function with a half-bandwidth of $0.5\,\omega_{eg}$. The dashed line is the exact EIT dispersion.}
\label{fig:polariton}
\end{center}
\end{figure*}

\subsection{Optical Bloch equations}

We assume the internal dynamics of the atoms to be restricted to three levels in the $\Lambda$ configuration sketched in Fig.\ref{fig:polariton}(a): the ground state $\ket{g}$, a long-living metastable state $\ket{m}$, radiatively decoupled from $\ket{g}$, and an optically active excited state $\ket{e}$, with a lifetime $\gamma_e^{-1}$. 
The lifetime of the $\ket{m}$ state is assumed to be much longer than any other time scale in the system: $\gamma_{m}^{-1} \gg (\gamma_e^{-1}, t_{max})$ where $t_{max}$ is the time of observation of the system. The atomic ensemble is assumed to be initially in the ground state. 

Two lasers are shone on the atoms: the control (or dressing) field is a strong laser beam coupled to the $\ket{m} \rightarrow \ket{e}$ transition of carrier frequency $\omega_c$ and (time-dependent) Rabi frequency $\Omega_c$. This field is considered to be an external field that can be controlled at will. 
The focus of the present work is centered on the propagation of another laser pulse of carrier frequency $\omega_p$ close to resonance with the $\ket{g}\rightarrow \ket{e}$ transition of dipole moment $d_{eg}$, and has an electric field
\begin{equation}
E(x,t)=\mathcal{E}(x,t)\,e^{-i\omega_pt}+\textrm{c.c.}.
\end{equation}
This probe field is assumed to be weak enough to be within the linear regime where a linearized version of the optical Bloch equations~\cite{CCT4,CPT_review} can be used. Within the well-known rotating-wave approximation, only quasiresonant terms for the atom-laser interaction are considered.

In order to eliminate the carrier frequency, we can rewrite the optical Bloch equations in terms of the reduced coherences $\tilde{\rho}_{eg} = \rho_{eg}\,e^{i \omega_pt}$ and $\tilde{\rho}_{mg} = \rho_{mg}\, e^{i (\omega_p - \omega_c)t}$:
\begin{subeqnarray}
\frac{\partial \tilde{\rho}_{eg}}{\partial t} & = & -\left(\frac{\gamma_e}{2} + i \delta_e \right) \tilde{\rho}_{eg} + i  \frac{d_{ge}\mc{E}}{\hbar} - i \frac{\Omega_c}{2} \tilde{\rho}_{mg},  \\
\frac{\partial \tilde{\rho}_{mg}}{\partial t} & = & -\left(\frac{\gamma_m}{2} + i \delta_R \right) \tilde{\rho}_{mg} - i \frac{\Omega_c}{2} \tilde{\rho}_{eg}.
\end{subeqnarray}
Here, $\delta_e = \omega_e - \omega_g - \omega_p$ is the one-photon detuning of the probe field from the $\ket{g}\rightarrow\ket{e}$ transition, while $\delta_R = \omega_m + \omega_c - \omega_g - \omega_p$ is the detuning of the probe and control beams from the so called Raman two-photon transition connecting the ground to metastable states.

As the transition $\ket{g}\rightarrow \ket{m}$ has a vanishing dipole moment and the coherence $\rho_{em}$ is of higher order in $\mc{E}$, the atomic polarization can be written in terms of the coherence $\rho_{eg}$ only:
\begin{equation}
\label{eq:pol}
P(x,t) =  n(x)\, d_{eg} \, \rho_{eg}(x,t) + \textrm{c.c.}
\end{equation}
Here, $n(x)$ is the atomic density.

\subsection{Modified slowly varying envelope approximation (MSVEA)}

In the present work we restrict our attention to a one-dimensional geometry at normal incidence. As the different polarizations of electromagnetic (e.m.) field are in this case decoupled, the vector nature of Maxwell equations disappears and one is left with a scalar problem for each component:
\begin{equation}
\left(\frac{\partial^2}{\partial x^2} - \frac{1}{c^2} \frac{\partial^2}{\partial t^2} \right) E(x,t) = \mu_0 \frac{\partial^2}{\partial t^2} P(x,t).
\label{eq:maxwell}
\end{equation}
Here $P(x,t)$ is the polarization (\ref{eq:pol}) of the atomic medium. The constants $c$ and $\mu_0$ are, respectively, the velocity of light in vacuum and its magnetic permeability.

We assume that the pulse envelope $\mc{E}(x,t)$ varies on a characteristic time scale much slower than the carrier frequency $\omega_p$.
Under this approximation, we can perform a \emph{modified slowly varying envelope approximation} (MSVEA) and neglect the second-order time derivatives of the envelope. 
Assuming the same form $P(x,t) = \mc{P}(x,t) e^{- i \omega_p t} + \textrm{c.c.}$ for the atomic polarization, the Maxwell equation (\ref{eq:maxwell}) can then be written in the form
\begin{equation}
\left[\frac{\partial^2}{\partial x^2} + \frac{\omega_p}{c^2}\left( \omega_p + 2i \frac{\partial}{\partial t} \right)\right]\mc{E} = - \mu_0\, \omega_p^2\, \mc{P}.
\label{eq:SVEAMaxwell}
\end{equation}
Here we have neglected the first-order derivative in time of the polarization envelope which is proportional to second-order time derivative of the electric field \cite{EIT_dispersion}.

It is important to note that, differently from the conventional SVEA discussed in textbooks (e.g., \cite{scully}), 
we have not separated out the spatial part of the envelope from the carrier and we have retained all the derivatives of the field with respect to the spatial coordinates. This crucial feature of the MSVEA equation (\ref{eq:SVEAMaxwell}) is required when one investigates configurations involving abrupt jumps in the spatial distribution of atoms $n(x)$.

The optical polarization (\ref{eq:pol}) arising from the optical Bloch equations has to be plugged back into the Maxwell equation (\ref{eq:SVEAMaxwell}).
After choosing an appropriate normalization, the resulting Maxwell-Bloch (MB) equations can be cast in the form
\begin{subeqnarray}
\label{eq:MB}
\frac{\partial\mc{E}}{\partial t} &=& \frac{i}{2} \left(\frac{\partial^2}{\partial x^2} + 1 \right)\mc{E} + i \sqrt{D} \tilde{\rho}_{eg}\slabel{eq:MB1},\\
\frac{\partial\tilde{\rho}_{eg}}{\partial t}& =& - \left(\frac{\gamma_e}{2} + i \delta_e \right)\tilde{\rho}_{eg} + i \sqrt{D} \mc{E} - i\frac{\Omega_c}{2} \tilde{\rho}_{mg}\slabel{eq:MB2},\\
\frac{\partial\tilde{\rho}_{mg}}{\partial t} &=& - \left(\frac{\gamma_m}{2} + i \delta_R \right)\tilde{\rho}_{mg} - i\frac{\Omega_c}{2}\tilde{\rho}_{eg}. \slabel{eq:MB3}
\end{subeqnarray}
In particular, $\omega_{p}$ is used as the unit for frequency, and $k_{p} = (\omega_{p}/c)$ for the wave vector. 
The electric field is measured in terms of $\mc{E}_0 = \sqrt{n \hbar\omega_{p}/2\epsilon_0}$. 
The physical meaning of this choice is related to the energy density in the system: the energy density associated to the atoms is $W_{at} = n \hbar \omega_{p} |\tilde{\rho}_{eg}|^2$, while the energy in the e.m. field is \cite{Landau} $W_{e.m.} = 2\epsilon_0 \mc{E}_0^2 |\mc{E}|^2$: $\mc{E}_0$ is then the electric field associated to a polariton in which the excitation is exactly shared between atoms in the excited state and photons, $|\mc{E}|^2 = |\tilde{\rho}_{eg}|^2=1, W_{at} = W_{e.m.}$. \\
The strength of the light-matter coupling is quantified by the adimensional parameter
\begin{equation}
D = \frac{d_{ge}^2n}{2\epsilon_0\hbar\omega_p}.
\label{eq:rabisplitting}
\end{equation}
In the case of a homogeneous system of $N$ atoms, the $D$ coefficient is straightforwardly related to the atom-field coupling constant $g$ as defined in \cite{EIT_dark_polariton},
\begin{equation}
D\omega_p^2 = g^2N.
\end{equation}

\subsection{Features and limitations of MSVEA}

Before proceeding, it is important to assess the features and limitations of the MSVEA approach that we introduced in the previous section. This approximation leads, in fact, to Eqs. (\ref{eq:SVEAMaxwell}) and (\ref{eq:MB}) that differ from the standard formalism used for EIT-related problems and offer important advantages for the specific problems under consideration here.

In the absence of atoms the MSVEA equation (\ref{eq:SVEAMaxwell}) with $\mc{P}=0$ gives the following approximate dispersion for the free e.m. field:
\begin{equation}
 \omega(k) = \frac{c^2k^2 + \omega_p^2}{2\omega_p}.
\label{eq:omega_k}
\end{equation}
On one hand, this dispersion is able to simultaneously describe both the forward ($k>0$) and the backward ($k<0$) propagating photons. This will be useful to handle reflectivity problems without the need for a coupled mode theory. In particular, this approximate dispersion correctly reproduces the 
exact one in the neighborhood of $k=\pm k_p$, for example
\begin{subeqnarray}
  \label{eq:res_disp}
  \omega(\pm k_p) &=& c |k_p|, \\
  \frac{d\omega}{dk}\bigg{|}_{\pm k_p} &=& \pm c.
\end{subeqnarray}
On the other hand, the deviation from the linear dispersion of light that is due to the curvature of the approximate dispersion (\ref{eq:omega_k}) is responsible for a spurious wave-packet broadening. However, this effect start to be important over propagation lengths that are much longer than the ones under investigation here.

At the interface with a generic semi-infinite medium of linear susceptibility $\chi(\omega)$, the reflectivity of a monochromatic wave at normal incidence can be straightforwardly calculated from (\ref{eq:SVEAMaxwell}) as
\begin{equation}
r(\omega)=\frac{1-k'/k}{1+k'/k}
\label{eq:refl}
\end{equation}
where the MSVEA wave vectors in vacuum and in the medium are respectively given by the dispersion laws
\begin{eqnarray} 
c k &=& \omega_p\,\sqrt{1 + 2(\omega - \omega_p)/\omega_p}, \\
c k'&=& \omega_p\,\sqrt{1 + \chi(\omega) + 2(\omega - \omega_p)/\omega_p}.
\end{eqnarray} 
Provided the frequency $\omega$ is close to the carrier $\omega_p$, the approximate reflectivity (\ref{eq:refl}) is accurate up to corrections of the order $(\omega-\omega_p)/\omega_p$.
This condition is well satisfied in an EIT medium in the frequency region around resonance as the light propagation is dominated by the frequency dispersion of the susceptibility $\chi(\omega)$.

To conclude this section, it is important to note that at the level of the MSVEA approximation one is allowed to replace the $c^2k^2$ term in (\ref{eq:omega_k}) with a generic function $f(k)$ that satisfies the conditions (\ref{eq:res_disp}).
This feature is of great interest when one is to numerically solve the set of Eqs. (\ref{eq:MB}).
In fact, a proper choice of $f(k)$ makes it possible to significantly increase the time step of the simulation~\cite{numerical}: in the remainder of the article, we will use the Erf-shaped form
\begin{equation}
  f(k) = \omega_p^2 \left[ 1 + \textrm{Erf}\left(\sqrt{\pi}\frac{|k| - k_p}{k_p} \right)\right]
\end{equation}
which results in a bandwidth of the order of $\omega_p$, wide enough to avoid introducing spurious physics in the frequency region of interest.
The resulting dispersion for the EIT system is shown in Fig.\ref{fig:polariton}, which is in excellent agreement with the exact one in the frequency region of interest. In particular, the choice of a linear $f(k)$ in the vicinity of $\omega_p$ allows to suppress the effect of the spurious dispersion of the wave packet that would be otherwise introduced by the mSVEA. 
We have also checked that the numerical results presented in the remainder of the article do not depend on the specific choice of $f(k)$.

\section{Polaritons in a homogeneous and static EIT medium}

Before entering in the discussion of the complex features of the propagation in spatially inhomogeneous and time-dependent media, it is useful to briefly review the main features of light propagation in the simplest case of homogeneous and static EIT medium.

At the level of linear optics considered in the present article, the response of a stationary medium is summarized in its susceptibility $\chi(\omega)$. In the case of a EIT medium, this is straightforwardly calculated from the Bloch equations [(\ref{eq:MB2}) and (\ref{eq:MB3})] to be
\begin{equation}
\chi (\omega_p) = 2D\omega_p\left(\delta_e - i \frac{\gamma_e}{2} - \frac{(\Omega_c/2)^2}{\delta_R - i(\gamma_m/2)}\right)^{-1}.
\label{eq:EITsusceptibility}
\end{equation}
Once plugged into the Maxwell equation, this form of $\chi(\omega)$ gives the three polariton branches \cite{EIT_dark_polariton, carusotto_casymir} that are shown in Fig. \ref{fig:polariton}(b). In this figure, as well as in the rest of the article we focus our attention on the resonant case $\omega_c=\omega_e-\omega_m$. The width of the Rabi splitting between the polariton bands at resonance is fixed by the light-matter coupling to $(2 \sqrt{D} \omega_p)$.
Note how the Erf-shaped form of $f(k)$ chose for the calculation does not affect the polariton dispersion in the region of interest.

Throughout this article, we are mostly interested in the central band, the so-called middle polariton or dark polariton (DP) \cite{EIT_dark_polariton}. The width of this DP branch is proportional to the dressing Rabi frequency $\Omega_c$. Exactly on Raman resonance $\delta_e=0$, the group velocity is given by
\begin{equation}
v_{gr} = \frac{c} {\displaystyle{1 + \frac{\omega_p}{2}\frac{\partial \chi}{\partial \omega_p}}}
= \frac{c}{ \displaystyle{ 1 + \frac{D \omega_p^2}{(\Omega_c/2)^2}}}.
\label{eq:MPvgr}
\end{equation}

In contrast to standard optical media where an optical resonance is generally associated to a high interface reflectivity and a strong absorption~\cite{jackson,bariani,EIT_reflectivity,atomic_resonance},
three-level systems under a coherent dressing allow for a very slow group velocity, a negligible interface reflectivity, and negligible absorption~\cite{EIT_review}.
This striking effect is due to the vanishing value of the susceptibility $\chi(\omega)$ exactly on resonance combined with a strong value of its frequency dispersion. Spontaneous emission from the excited state is suppressed thanks to the destructive interference between the different excitation paths, while dephasing between the ground and metastable states is generally very small in cold atom systems, $\gamma_m\simeq 0$.

For a polariton wave at a wave vector $k$, the group velocity and the lifetime are in fact related to the relative weights of the radiation and matter excitation components that are obtained by diagonalizing~\cite{carusotto_casymir} the set of Eqs. (\ref{eq:MB}):
\begin{eqnarray}
v_{gr}(k) &=& c\; \frac{\left|\mc{E}(k)\right|^2}{\left|\mc{E}(k)\right|^2 + \left|\tilde{\rho}_{eg}(k)\right|^2 + \left|\tilde{\rho}_{mg}(k)\right|^2},
\label{eq:vgrratio} \\
\gamma(k) &=& \gamma_e \; \frac{\left|\tilde{\rho}_{eg}(k)\right|^2}{\left|\mc{E}(k)\right|^2 + \left|\tilde{\rho}_{eg}(k)\right|^2 + \left|\tilde{\rho}_{mg}(k)\right|^2}.
\label{eq:EITabsorption}
\end{eqnarray}
The prediction (\ref{eq:EITabsorption}) for the decay rate of the different bands is plotted in Fig.\ref{fig:polariton}(c). As expected, the decay rate is exactly vanishing on resonance (for $\gamma_m=0$) and grows quadratically in the wave vector $k$:
\begin{equation}
\gamma(k) = \frac{1}{2} \Imag{\frac{d^2 \omega(k)}{dk^2}\bigg|_{\delta_R=0}} (k - k_p)^2.
\label{eq:approxabsorp}
\end{equation}
To leading order, the second-order derivative of the EIT dispersion is then purely imaginary \cite{EIT_dispersion}:
\begin{equation}
\frac{d^2\omega}{dk^2}\bigg|_{\delta_R=0} = - i \frac{\gamma_e}{c} \frac{D \omega_p^2}{(\Omega_c/2)^4} v_{gr}^3.
\end{equation}
For a sample length $L$, this recovers the well-known \cite{EIT_review, EIT_dark_polariton} Gaussian transmittivity window of width 
\begin{equation}
\Delta\omega_{TR} = \frac{1}{\sqrt{2}} \frac{(\Omega_c/2)^2}{\sqrt{\gamma_e \;D\omega_p^2}} \sqrt{\frac{c}{L}}.
\end{equation}

\section{Multilayer system: the EIT chain}
%Idea: a layered EIT medium

\begin{figure}[htbp]
\begin{center}
\includegraphics[width = 0.9\columnwidth,clip]{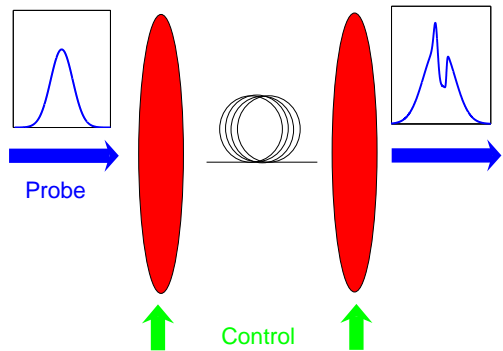}
\caption{Pictorial scheme of a double-layer EIT chain. The shown configuration with orthogonal probe and control beams~\cite{EIT_experiment} allows for independent modulation of $\Omega_c$ in the different layers, but the predictions of the present article are straightforwardly extended to other geometries.}
\label{fig:EITchain}
\end{center}
\end{figure}

We consider a pulse of light launched into a layered geometry consisting of several atomic EIT media separated by empty regions of space. A pictorial view of the layered system under consideration -the \emph{EIT chain}- is shown in Fig.\ref{fig:EITchain}.
We assume the atoms in the different EIT layers to have the same Raman frequency. The probe pulse carrier is taken exactly on Raman resonance and the pulse bandwidth is assumed to fit within the EIT frequency window. The propagation of the pulse across the system can be simulated using the MB formalism (\ref{eq:MB}) described in the previous sections with a spatially dependent $D(x)$ and $\Omega_c(x)$. 

Two cases can be distinguished: a {\em static} case where all the parameters describing radiation-matter interaction remain costant in time, and a \emph{dynamic} configuration when some parameter is varied in time while the pulse is propagating across the system. 
In what follows, we will concentrate our attention on a time modulation of the control field amplitude $\Omega_c$. Other dynamic schemes may involve a modulation of the atomic resonance frequency and/or of the dressing field frequency~\cite{bariani,EIT_spatial}.

% EIT chain scheme
% Solving numerically the MB equations

% Static propagation
\subsection{Static case}

In a static situation, the pulse propagates across the whole multilayer structure with negligible distortion. The propagation time is equal to the sum of the thickness of each layer divided by the corresponding group velocity. 

While the spatial shape of the emerging pulse remains unchanged with respect to the incident pulse, the shape inside the structure is significantly modified. As shown in Fig.\ref{fig:simple_modulation}(a), a pulse entering an EIT layer is in fact spatially compressed by $v_{gr}/c$ as a consequence of the reduced group velocity and a reversed process takes place when the pulse leaves the layer, which restores the initial shape. 
Correspondingly, the envelope of the electric field remains continuous at all interfaces, but its derivative has discontinuities proportional to the group velocity mismatch between neighboring layers.

An important distinction has therefore to be carefully made in the notation: the length of the wave packet within the EIT layer will be denoted by $\bar{\sigma}_x = \sigma_t\,v_{gr}$ while its length in vacuum will be denoted by $\sigma_x = \sigma_t \, c$.

% Dynamical modulation 
\subsection{Dynamic case}

The easy tunability of the properties of the dressing field together with the slow propagation of the DP allow for an efficient dynamic modulation of the propagating pulse. Taking advantage of spatial inhomogeneities, the dynamic EIT chain can be the paradigm for a new class of dynamic photonic structures \cite{DPS,EIT_spatial} which are able to perform a simultaneous spatial and temporal modulation of the wave-packet profile.
The mechanism of the pulse modulation in complex geometries can be understood in terms of two basic building blocks: the homogeneous layer, when the pulse is completely contained in a single homogeneous EIT layer during the whole modulation sequence, and the interface, when the modulation takes place while the pulse is overlapping two neighboring layers. 

\subsubsection{Homogeneous layer}

When the whole pulse fits into a homogenous EIT layer, its dynamics can be easily understood within the polariton picture discussed above and it is fully determined by the conservation of wave vector $k$.

In the limit of a very slow modulation, the pulse adiabatically follows the evolution of the DP state \cite{EIT_dark_polariton} and propagates at the instantaneous group velocity with no change in its length. The peak electric field follows the magnitude of the photonic component (\ref{eq:vgrratio}) and is determined by the ratio of the final to initial group velocity $v_{gr}^{f}/v_{gr}^{i}$. In particular, it does not depend on the temporal shape of the modulation. Examples of such modulations are illustrated in Fig. \ref{fig:simple_modulation}(b).
In the case in which the group velocity modulation is brought back to the initial value $v_{gr}^{f} = v_{gr}^i$, the pulse emerges with an unchanged profile and the layer can be considered as a very compact, yet programmable delay line.
It is interesting to note that the dynamical modulation of the polariton branches ensures that, if the pulse fits into the EIT window at the entrance, then it will fit into it at all later times~\cite{DPS,EIT_spatial}.

The main effect of a finite ramp time $\tau$ is to couple the DP to the upper and lower polariton branches at the same $k$. The effective matrix element of this coupling is proportional to the modulation rate of the control field amplitude $d\Omega_c/dt$. This matrix element is to be compared to the frequency splitting between the bands, which leads to the following quantitative criterion for adiabaticity~\cite{Messiah,EIT_adiabatic,FB_PhD_thesis}
\begin{equation}
\left|\frac{d\Omega_c}{dt}\right| \ll D \omega_p^2
\label{eq:adiab}
\end{equation}

% interface: reshaping, making cuts along the pulse
\subsubsection{Interface}

\begin{figure}[htbp]
\begin{center}
\includegraphics[width=0.9\columnwidth]{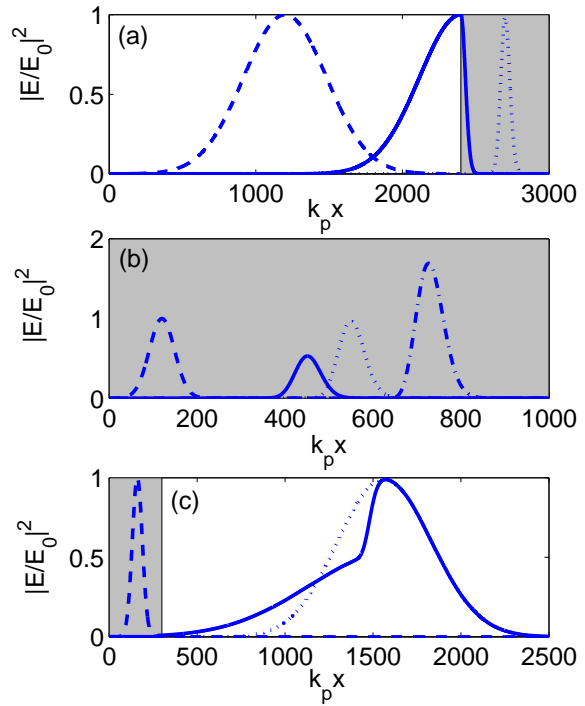}
\caption{Examples of modulation of a pulse via a space- and time-dependent EIT. The propagation of the pulse is calculated by solving the MB equations (\ref{eq:MB}). In all of the panels, the pulse moves from left to right and the shaded regions correspond to a EIT medium. The dashed curves show the initial pulses. (a) Propagation across a static interface. Solid line, pulse shape while being spatially compressed in entering into a medium with $v_{gr}^i = 0.11\;c$; Dotted line, pulse shape once it has completely entered the EIT medium. (b) Wave-packet propagation in a homogeneous EIT layer. Dotted line, propagation without modulation; solid line, result of a slow down ramp with $v_{gr}^f=v_{gr}^i/2$; dot-dashed line, result of a speed-up ramp $v_{gr}^f=1.8\,v_{gr}^i$. (c) Exit from a EIT medium into vacuum. Dotted line, propagation without modulation; solid line, result of a slow-down ramp with $v_{gr}^f=v_{gr}^i/2$. 
All the panels have been calculated for a pulse carrier exactly on Raman resonance $\delta_e=0$, a ramp time $\omega_p\tau=100$, and a pulse width $\omega_p\sigma_t=400$. Material parameters: $D = 0.01$, $\gamma_e = 10^{-3}\omega_p$.}
\label{fig:simple_modulation}
\end{center}
\end{figure}

A dynamical modulation taking place while the pulse overlaps an interface provides a simplest way of reshaping the pulse: only the part of the pulse which is located inside the EIT layer is in fact affected by the modulation of the dressing field. 
In contrast to the the spatially homogeneous case considered earlier in this article, the shape of the emerging pulse now strongly depends on the details of the modulation ramp even in the adiabatic limit. 
This crucial fact is illustrated in Fig.\ref{fig:simple_modulation}(c): the group velocity of a EIT medium is reduced while a pulse is exiting into vacuum.

The results can be understood by isolating three portions of the pulse: the first part is already in vacuum when the modulation begins, while the last part is still in the EIT medium when the modulation is completed.
 
The first part is therefore not affected by the modulation, while the electric field amplitude of the third part is homogeneously lowered. 
When also this part of the pulse eventually exits into vacuum, its spatial length is stretched out by a factor $c/v_{gr}^f$. As compared to the front of the pulse, the stretching is larger by a factor $v_{gr}^i/v_{gr}^f$, which results into the strongly asymmetric pulse shape that is visible in Fig.\ref{fig:simple_modulation}(c).

Finally, the modification of the central part of the pulse depends in a nontrivial way on the details of the ramp. For a fast [but still adiabatic as compared to the interband splitting, according to (\ref{eq:adiab})] modulation, the first and third parts of the pulse are connected by a sharp jump in the electric field amplitude. For slower ramps, this jump is replaced by a smoother crossover.

\subsubsection{Defect}

The physics of a defect geometry can be understood along these same lines. 
Two regions of a homogeneous medium are separated by a thin layer with a different group velocity. The thickness $L_{d}$ of the defect region is taken to be small as compared to the effective length of the pulse in this layer. 
For the sake of simplicity, we restrict our attention to the simplest situations of a EIT medium with a  vacuum defect and of a single EIT slab in vacuum. 

% Figure: defect modulation
\begin{figure}[htbp]
\begin{center}
\subfigure{\includegraphics[width=0.9\columnwidth]{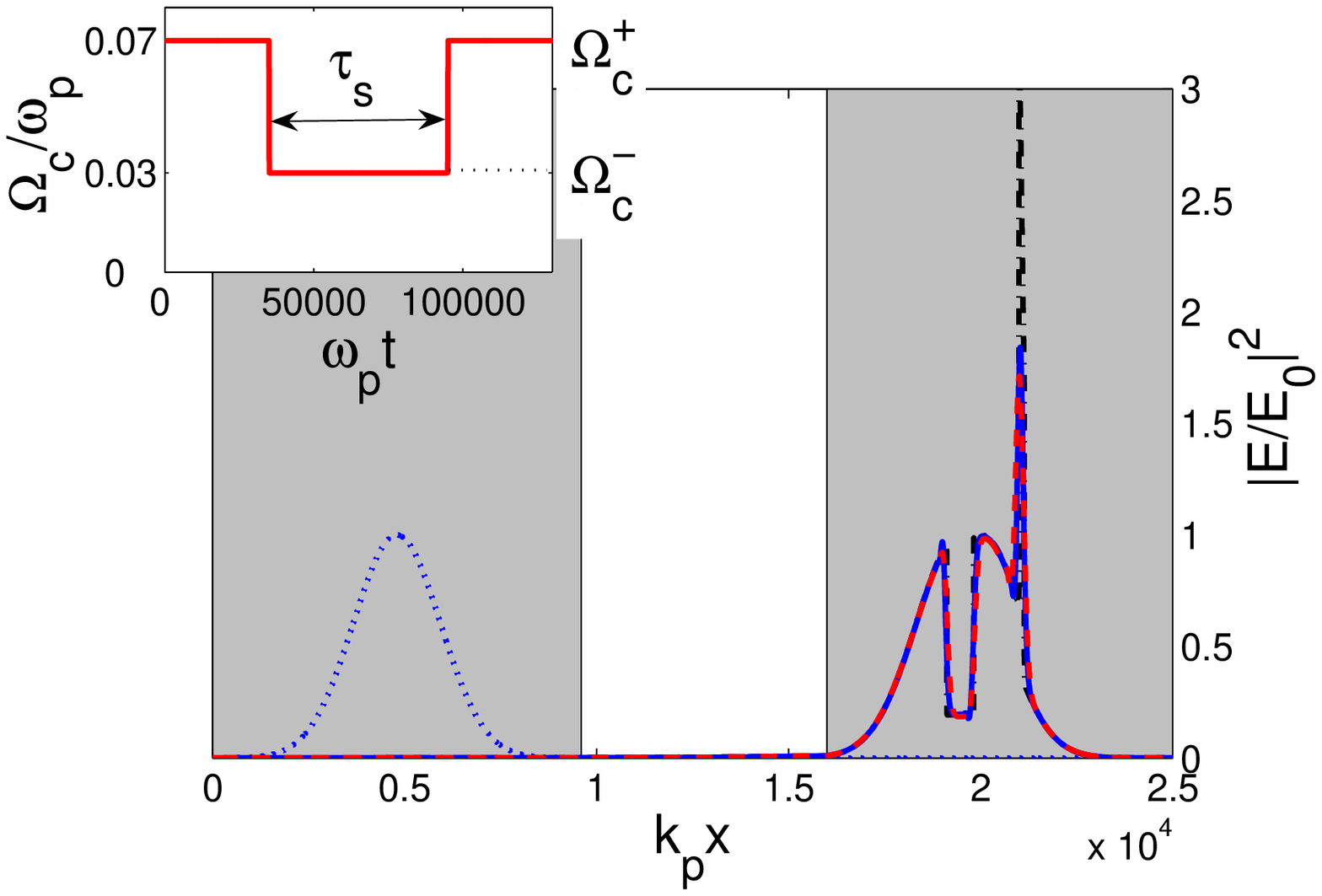}}
\subfigure{\includegraphics[width=0.9\columnwidth]{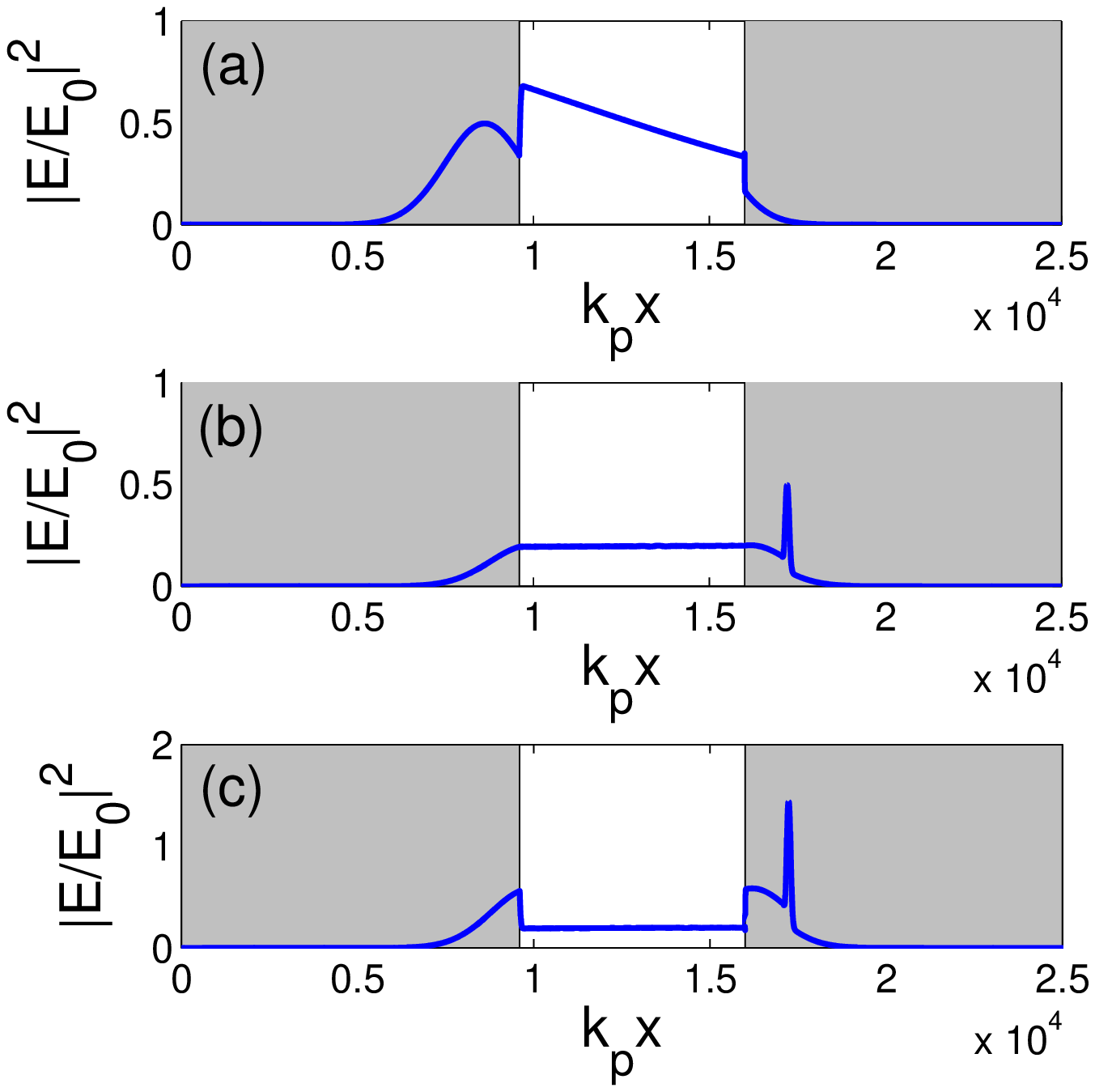}}
\caption{Modulation of a wave packet using a vacuum defect. In the EIT medium, the group velocity is decreased from $v^+_{gr} = 0.11\,c$ to $v^-_{gr} = 0.02\,c$ and then increased back to $v^+_{gr}$ as shown in the inset. (Main panel) Pulse at the beginning (dotted blue line) and at the end of the process (solid blue line); comparison with the results of the continuity equation for polariton flow (\ref{eq:continuityelectric}) (black dot-dashed line) and the numerical solution of the effective equation (\ref{eq:effectiveequation}) (red dashed line). (Inset) Temporal dependence of $\Omega_c$. (Bottom panels) Three snapshots during the propagation time. (a) Slow-down ramp; (b) storage time; (c) speed-up ramp. 
Parameters: $\gamma_e = 10^{-3}\omega_p$, pulse length $k_p\bar{\sigma}_x = 1600$, defect thickness $k_p L_d = 6400$, ramp time $\omega_p\tau=100$, storage time $\omega_p\tau_s=60000$.}
\label{fig:defect_modulation}
\end{center}
\end{figure}

The case of a vacuum defect is illustrated in Fig.\ref{fig:defect_modulation}. In particular, we consider the long pulse limit $\sigma_x \gg L_{d}$ in which the pulse amplitude can be considered as almost homogeneous across the defect.
The temporal shape of the modulation of $\Omega_c$ is shown in the inset: its time scale is assumed to be fast as compared to the pulse duration, but still slow as compared to the interband adiabaticity requirement (\ref{eq:adiab}). $v_{gr}^\pm$ are the maximum and minimum values of the group velocity; $\tau_s$ is the interval between the two ramps, that is, the {\em storage} time.  

The modulation takes place in three stages: the slow-down ramp, the storage time and the speed-up ramp.  
When $\Omega_c$ is first decreased, the pulse intensity in the atomic medium is correspondingly reduced by a factor $v_{gr}^-/v_{gr}^+$, while the amplitude of the central part remains unchanged as it is sitting in vacuum [Fig.\ref{fig:defect_modulation}(a)]. 
When this part of the pulse re-enters the EIT medium, it results in a spatially compressed, narrower peak of width $L_{d} \, v_{gr}^-/c$ [Fig.\ref{fig:defect_modulation}(b)]. 
The part of the pulse that crosses the defect during the storage time does not experience any distortion: in Fig.\ref{fig:defect_modulation}(b) this part lies just behind the narrow peak and is $\tau_s v_{gr}^- - L_{d}\,v_{gr}^-/c$ long. 
The final speed-up ramp which restores the group velocity to its initial value $v_{gr}^+$ is responsible for an increase of the electric field amplitude in the EIT medium. This eventually results in a hole being left imprinted in the pulse profile correspondingly to the vacuum layer [Fig.\ref{fig:defect_modulation}(c)]. Once it has re-entered into the EIT medium, the length of this hole is equal to $L_{d}\,v_{gr}^+/c$.

This vacuum defect geometry is then the simplest example of a nontrivial modulation of the wave-packet profile: note that the resulting profile is very different from the one obtained in a homogeneous geometry, where the second ramp would simply compensate the first one.

% Figure: defect modulation
\begin{figure}[htbp]
\begin{center}
\includegraphics[width=0.9\columnwidth]{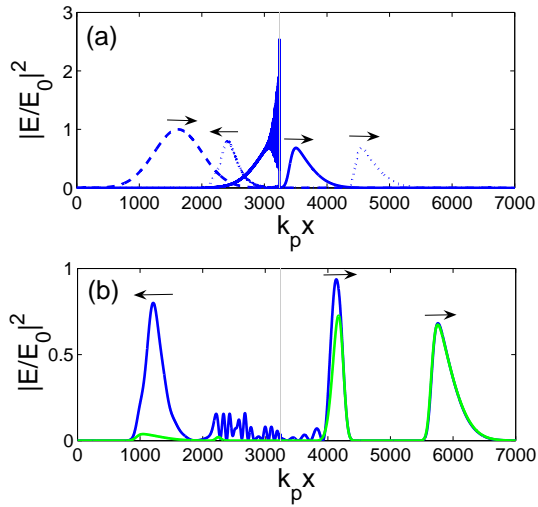}
\caption{Snapshots of the pulse profile during a light-storage process in a single EIT layer. $\Omega_c$ is modulated in time from $\Omega_c^+ = 0.07\, \omega_p$ to $\Omega_c^- = 0$ with the same shape as shown in the inset of Fig.\ref{fig:defect_modulation}. The storage time is $\omega_p \tau_s = 1350$. The initial group velocity in the EIT layer is $v_{gr} = 0.11\;c$. The pulse has a Gaussian shape with $k_p \sigma_x = 540$ in vacuum and the defect has a length $k_p L_d = 10$. (a) Initial pulse (dashed line), pulse hitting the defect and splitting immediately after the stopping ramp (solid line), and counterpropagating wave packets during the storage time (dotted line). (b) Emerging wave packets after the retrieval ramp. The arrows indicate the propagation directions of each pulse. Blue lines in panels (a,b) are in the absence of spontaneous emission $\gamma_e=0$. The green line in panel (b) is for $\gamma_e=0.07\,\omega_p$.}
\label{fig:storage}
\end{center}
\end{figure}

% EIT defect: light storage
The opposite case of an EIT layer in vacuum is presently of great experimental interest for light-storage purposes \cite{EIT_dark_polariton,optimal_light_storage}. 
The idea is very simple: by switching off the control field $\Omega_c$ while the wave packet is inside the EIT medium, the DP is fully mapped into a metastable coherence $\rho_{mg}$. As this has a vanishing group velocity, it can remain stored in the atoms for macroscopically long times. 
When the control field $\Omega_c$ is switched on again, the wave packet is retrieved. 
The main limitations to the efficiency of a light-storage process originate from spontaneous emission processes from the excited state (a finite $\rho_{eg}$ component is always present for any pulse of finite duration), leakages due to the finite optical depth of sample as the usually considered systems are shorter than the effective length of the pulse, and ground-state decoherence $\gamma_m>0$~\cite{EIT_dark_polariton,optimal_light_storage}. 
Even though this last effect sets the ultimate limit to the performances of light-storage experiments, for the parameters considered in the present work its effect is negligible as compared to the other processes.

A situation similar to the experimental realization in \cite{EIT_bloch} is simulated in Fig.\ref{fig:storage}: because of the limitations in the numerical solution of the MB equations, we have been forced to consider a EIT medium with a much bigger $v_{gr}/c$. Anyway, a suitable rescaling of all other parameters makes it possible to observe the same dynamics. 

During the storage time, $\Omega_c$ is brought to zero with the same temporal profile as shown in the inset in Fig.\ref{fig:defect_modulation}: again, this process consists of three stages.  During the stopping ramp [Fig.\ref{fig:storage}(a)], the signal is cut into three parts. The front part, which has already crossed the defect, is not affected by the modulation and keeps on propagating almost undisturbed. The central slice that was contained in the EIT layer at the stopping time remains coherently stored in the medium as an atomic polarization. 
The back part of the pulse hits the medium when this is no longer optically dressed and its fate strongly depends on the spontaneous emission rate from the $\ket{e}$ state.

If spontaneous emission is negligible, this part of the pulse is reflected back albeit in a strongly distorted way~\cite{EIT_reflectivity,bariani,EIT_dark_polariton}: this is for example, the case of a Mott-insulator state of two-level atoms where spontaneous emission is suppressed as a consequence of the ordered lattice structure \cite{hopfield_model} and information on the interface structure can be inferred from the shape of the reflected wave packet.
On the other hand, as shown by the green (light gray) line in Fig.\ref{fig:storage}(b), the reflected wave packet is almost completely absorbed in the presence of a significant spontaneous emission rate. 

When the retrieval ramp is finally applied, the excitations stored in the EIT layer are free to propagate out of the atomic medium. The efficiency of this retrieval process (defined as the ratio of the intensity of the retrieved pulse to the intensity of the incident pulse) is around $15 \%$ for this simulation, which qualitatively agrees with the estimation in \cite{EIT_bloch}. As shown in Fig.\ref{fig:storage}(b), this retrieved wave packet is weakly affected by spontaneous emission. 

\section{Polariton flow: effective description}

The MB formalism (\ref{eq:MB}) gives a complete picture of the pulse propagation which is able to take into account interband transitions as well as reflection at the interfaces.  As the solution of the three coupled equations is time- and memory-consuming, it quickly becomes unfeasible for growing values of the velocity mismatch between the different media.

For this reason, an effective approach able to investigate the ultraslow light regime can be of great interest. 
As the dynamic modulation of the pulse does not result into an increased reflection as compared to the static case, the propagation dynamics can be described in terms of a single continuity equation for the polariton flow. In particular, both absorption and the spatial inhomogeneity of the structure are fully included in this model equation.

\subsection{Continuity equation}

In terms of the polariton density $n_p(x,t)$, the continuity equation describing conservation of the total number of polaritons reads
\begin{equation}
\frac{\partial }{\partial t} n_p(x,t) + \frac{\partial }{\partial x} \left(n_p(x,t) v_{gr}(x,t)\right) = 0.
\label{eq:continuity}
\end{equation}
In the static situation ($v_{gr}(x,t) = v_{gr}^0(x)$), also the local polariton flux $n_p(x,t) v_{gr}^0(x)$ remains constant. It is useful to rewrite this equation in terms of the electric field intensity $I = |\mc{E}|^2$ corresponding to the polariton flux instead of the polariton density. For static multilayer geometries which are carachterized by abrupt changes in the polariton velocity, the polariton density shows, in fact, discontinuities while the electric field remains continuous, even at the interfaces. Taking into account the photonic weight and the group velocity (\ref{eq:vgrratio}), Eq.(\ref{eq:continuity}) then becomes
\begin{equation}
\frac{\partial I}{\partial t} + v_{gr} \frac{\partial I}{\partial x} =  - \frac{I}{v_{gr}}\frac{\partial v_{gr}}{\partial t}.
\label{eq:continuityelectric}
\end{equation}
The left-hand side of the equation contains the propagation terms for the static situation. The general solution is a mixed translation and dilation of the starting pulse $I^0(x)$ according to the trajectories in space-time which are the solution of the Cauchy problem: ($\dot{\xi} = v_{gr}(\xi), \xi(t) = x$) \cite{EIT_dark_polariton,EIT_shaping}. The specific solution clearly depends on the geometry. The right-hand side of (\ref{eq:continuityelectric}) is instead responsible for the amplitude variation in the dynamic case.

As a first example, this continuity equation model can be applied to the defect geometry of Fig.\ref{fig:defect_modulation},
\begin{equation}
v_{gr}(x,t) = v(t) \left(\theta(- x) + \theta(x - L_{d})\right) + c \theta(x)\theta(L_{d} - x),
\end{equation}
where the beginning of the defect is located at $x_{d} = 0$. 
For times longer than $t = (L_{d}/c)$, the solution for the electric field intensity is 
\begin{equation}
\label{eq:continuitydef}
I(x,t) = \left\{
 \begin{array}{ll}
  \displaystyle I^0\left( x - \mc{I}_0^t \right) \frac{v(t)}{v(0)} & x <  0\\
  \displaystyle I^0\left(-\mc{I}_0^{t-x/c}\right)  \frac{v(t - x/c)}{v(0)} & 0 <  x < L_{d} \\
  \displaystyle I^0\left(-\mc{I}_0^{t_0}\right)  \frac{v(t_0)}{v(0)} \frac{v(t)}{v(t_d)} & L_{d} <  x < L_{d} +  \mc{I}_{L_{d}/c}^t \\
  \displaystyle I^0\left(L_{d} - ct_d\right)  \frac{v(t)}{v(t_d)} & L_{d} + \mc{I}_{L_{d}/c}^t <  x < L_{d} + \mc{I}_{0}^t \\
  \displaystyle I^0\left(x - \mc{I}_0^t \right) \frac{v(t)}{v(0)} &  x >  L_{d} + \mc{I}_0^t 
 \end{array}
\right.
\end{equation}
Here $t_d(x)$ and $t_0(x)$ are the instants of time at which the point of the wave packet which is located at $x$ at the time $t$ has passed through $x = L_{d}$ and $x = 0$, respectively. They can be found by the conditions
\begin{subeqnarray}
&&x = L_{d} + \mc{I}_{t_d}^t, \\
&&t_0 = t_d - L_{d}/c.
\end{subeqnarray}
where we have defined 
\begin{equation}
\mc{I}_a^b = \int_a^b v(t') dt'.
\end{equation}
This analytic solution gives the black dash-dotted curve in the main panel of Fig.\ref{fig:defect_modulation}.

\subsection{Effect of losses}
Even if the carrier frequency $\omega_p$ sits exactly on Raman resonance, the finite time duration of the wave packet requires including absorption for the tails of the wave-vector spectrum \cite{EIT_relaxation}. This leads to a finite and momentum-dependent decay rate for the polaritons according to (\ref{eq:EITabsorption}). Taking inspiration from the approximated form  (\ref{eq:approxabsorp}) of the decay rate, a simple diffusion term can be used to model losses \cite{effective_model}.
The propagation Eq. (\ref{eq:continuityelectric}) for the intensity $I$ then becomes
\begin{equation}
\frac{\partial I}{\partial t} + v_{gr} \frac{\partial I}{\partial x} =  - \frac{I}{v_{gr}}\frac{\partial v_{gr}}{\partial t} + \frac{\partial}{\partial x}  \mc{D} \frac{\partial I}{\partial x},
\label{eq:effectiveequation}
\end{equation}
where
\begin{equation}
\mc{D} = i\,\left(\frac{d^2\omega}{dk^2}\right)_{\delta_R = 0}=v_{gr}\,\frac{c \gamma_e}{D \omega_p^2}
\end{equation}
is the diffusion coefficient. 
As shown in the main panel of Fig. \ref{fig:defect_modulation}, this effective model (red dashed line) captures all the features of the full MB calculation (blue solid line) with a good accuracy \cite{numerical2}. In particular, the effective model is able to perfectly reproduce the height of the peak that was instead overestimated by the simple diffusionless continuity equation (\ref{eq:continuityelectric}) (black dash-dotted line).

\section{Manipulation Schemes in Cold Gases}

The EIT chain introduced and discussed in the previous section can be implemented experimentally using clouds of ultracold atoms as the EIT media. Optical fibers can be used to fix the optical distance between the EIT layers \cite{delay_guide}. Other possible material realizations involve solid-state materials \cite{EIT_review}.

Realistic values for the system parameters can be obtained from \cite{EIT_experiment,steck}.
As a typical example, we consider a cloud of Na atoms of density $n = 8 \times 10^{19}\, \textrm{m}^{-3}$. For the optical transition, we use the $D_2$ line. As the ground state, we use the $\ket{g} = |3S_{1/2}, F = 1, m_F = -1\rangle$ sublevel; as the metastable state, we can use $\ket{m} = |3S_{1/2}, F = 2, m_F = -2\rangle$, and as the excited state we can use $\ket{e} = |3P_{3/2}, F = 2, m_F = -2\rangle$. In this case, we have $\omega_{eg} = (2\pi)508$ THz, $d_{eg} = 1.5 \times10^{-29}\textrm{C}{m}$, $\gamma_e = (2\pi)10$ MHz and $D = 3\times 10^{-9}$. 
For a control field of Rabi frequency $\Omega_c = (2\pi)17$ MHz, a group velocity $v_{gr} = 10^{-7} c$ is obtained along with an absorption coefficient $\mc{D} = 6\times10^{-7}\,\omega_{eg}/k^2_{eg}$.  The parameters of the recent experiment \cite{EIT_bloch} are not much different. 
From here forward, physical units are used in the figures.

We have used the effective Eq. (\ref{eq:effectiveequation}) to simulate several simple geometries. We first address the single-layer case and then we use the results to investigate more complicated multilayer structures.

\subsection{Single layer}
% Time delay along the slow light medium
We consider a single EIT layer where we inject a Gaussian pulse. As discussed in Sec. IV, the effect of the modulation depends on several parameters: a crucial quantity is the ratio $R = L/ \bar{\sigma}_x$ between the layer thickness and the pulse length. Here, we focus our attention on the $R< 1$ case and we restrict to the case in which the system size is much shorter than the absorption length
\begin{equation}
\ell_{abs} = \frac{v_{gr}}{2\gamma} = \left(\frac{v_{gr}}{c} \frac{D\omega_p^2}{\gamma_e}\sigma_t\right)\bar{\sigma}_x.
\end{equation}

The simplest quantity to measure in a static configuration is the time delay $L/v_{gr}$: for sufficiently small group velocities, this allows the detection of very small differences in $L$ due, for example, to defects in the structure. Clearly this measurement requires a good temporal resolution of the detector as well as some knowledge of the system parameters, in particular of $v_{gr}$.
 
The simplest example of dynamical modulation consists of a single-ramp modulation of $\Omega_c$. Only the small part of the pulse contained in the layer feels the modulation. As the layer is thin, the crossing time is often faster than the ramp time, which means that different parts of the pulse experience different portions of the ramp.
In this case, the effect of the modulation on each given slice of the pulse can be estimated as 
\begin{equation}
|\mc{E}^{f}|^2(T) = |\mc{E}^{i}|^2 \frac{v_{gr}^i + \Delta v_{gr}(T)}{v_{gr}^i},
\end{equation}
where $T$ is the time at which the slice exits the layer and the initial values $v_{gr}^i$ refer to the entrance of the slice in the EIT medium. 
If we approximate the ramp as linear, the variation in the electric field is then given by $(\Delta v/v_{gr}^i)(T/\tau)$, where $\Delta v$ is the amplitude of the group velocity ramp. 

\begin{figure}[htbp]
\begin{center}
\includegraphics[width = 0.9\columnwidth]{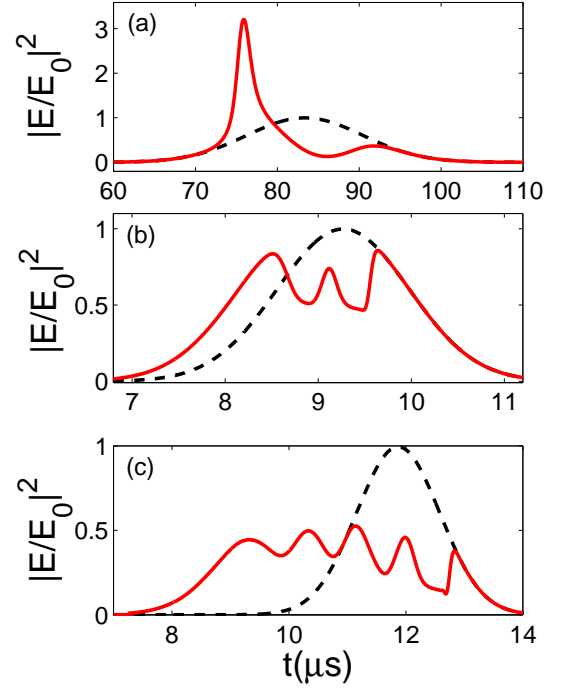}
\caption{Examples of wave-packet manipulation in several realizations of the EIT chain. All curves have been obtained using the model of Eq.(\ref{eq:effectiveequation}). Solid red lines show the modulated pulse, while dashed black lines result from static cases. (a) Single EIT layer [$L = 200 \mu$m, $v_{gr} = 10^{-7}\,c$, $\mc{D} = 6\times10^{-7}\,(\omega_p/k_p^2)$] with a Gaussian pulse ($\sigma_t = 10 \mu s$, $\ell_{abs} \approx 500 \bar{\sigma}_x$). Effect of a double ramp (as, e.g., in Fig.\ref{fig:defect_modulation}): $v_{gr}^+/v_{gr}^- = 10$, $\tau = 3.5\,\mu$s and storage time $\tau_s = 8\, \mu$s. (b) EIT double-layer structure [$L = 30 \mu$m (each layer), interlayer distance $\Delta L = 3\times10^{7} \mu$m, $v_{gr} = 5\times10^{-7}\,c$, $\mc{D} = 3\times10^{-6}\,\omega_p/k_p^2$] with a Gaussian pulse ($\sigma_t = 1 \mu$s, $\ell_{abs} \approx 400\,\bar{\sigma}_x$. Slow-down ramp in both layers: $\Delta v = -0.5\,v_{gr}$, $\tau = 50 ns$. (c) EIT chain with four layers [same as (b) but $\Delta L = 6\times10^{7} \mu$m] with Gaussian pulse [same as (b)]. Effect of a single slow-down ramp:  $\Delta v = -0.7\,v_{gr}$, $\tau = 50 ns$.}
\label{fig:modulation_schemes}
\end{center}
\end{figure}

As we have seen earlier in this article, a most interesting case consists of a double ramp, as illustrated in Fig. \ref{fig:modulation_schemes}(a): the pulse is slowed down and then accelerated back to the initial group velocity. In the $R < 1$ case under consideration here, the part of the pulse which is modulated during the slow-down ramp exits from the layer before the restoring ramp has begun. This latter ramp is then responsible for the creation of the peak in the trailing part of the pulse. The resulting shape is then very similar to the case shown in Fig.\ref{fig:defect_modulation}, yet time reversed. 
 
\subsection{Multilayer}
The single EIT layer is the basic building block for more complex geometries. As an example, the cases of single ramp in a double- and four-layer geometries are illustrated in Figs.\ref{fig:modulation_schemes}(b) and \ref{fig:modulation_schemes}(c). In particular, we restrict our attention to the case in which both the EIT and the vacuum layers are shorter than the pulse length.

The simultaneous modulation of the dressing field on both layers allows the creation of several similar structures on the same pulse, separated by a time depending on the distance between the layers. By choosing a fast-enough slow-down ramp, the peak that appears between the different layers can be shaped down to the absorption length. Its height can be enhanced by applying different ramps to the different layers, for example with opposite signs. The creation of such short and intense pulses can be of great interest in view of creating a strongly localized polariton, whose dynamics has been predicted to show peculiar features \cite{EIT_Dirac}. 

\section{Conclusion}
In conclusion, we have developed a Maxwell-Bloch formalism to investigate the propagation of a light pulse through a multilayer structure consisting of alternating layers of a generic EIT medium and vacuum. The formalism recovers the usual polariton dispersion in a homogeneous EIT medium and is able to include reflection at interfaces as well as absorption processes. 

The effect of a dynamical modulation of the control field amplitude on the pulse shape is studied and specific attention is paid to the case when the modulation takes place while the pulse is overlapping an interface: as reflection and losses remain negligible as long as the pulse fits in the EIT window, abrupt cuts on the wave-packet profile can be performed. 

In order to rapidly and efficiently simulate the propagation in geometry with strong velocity mismatch, an effective equation for the wave-packet propagation is developed. The resulting modified continuity equation is shown to account for both propagation and absorption effects. A good agreement with the full Maxwell-Bloch calculations is found. 

Starting from this equation several manipulation schemes have been proposed and characterized using realistic parameters for ultracold atomic clouds: contrary to standard light-storage techniques, our proposal is able to take full advantage of the reduced size of typical atomic clouds. A most interesting possibility is the creation of highly localized peaks. 

Our calculations confirm that the wave-packet manipulation can be performed in a coherent and almost lossless way, which opens interesting perspectives toward the quantum processing of single-photon wavefunctions for optical quantum computing applications. From a broader standpoint, the peculiar reflection properties at the interfaces of atomic media can be useful for polariton trapping and guiding as well as interface characterization.

\begin{acknowledgments}
F.B. aknowledges useful discussions with M. Fleischhauer, J. Otterbach, R. Unanyan and G. Nikoghosyan.
\end{acknowledgments}


\begin{thebibliography}{99}
\bibitem{numerical} We solve the set of MB equations (\ref{eq:MB}) by means of a finite-difference time-domain (FDTD) scheme with a fourth-order Runge-Kutta scheme to propagate the fields in time. Spatial derivatives of the fields are evaluated in momentum space using a discrete fast Fourier transform. 
The crucial issue limiting the speed of the numerical calculation is the time step of the FDTD evolution: its maximum value is set by the width of the frequency band considered in the problem. 
A proper choice of $f(k)$ with suitable upper and lower bounds makes it possible to restrict this bandwidth without affecting the physics.
\bibitem{effective_model} We stress that this diffusive model is able to reproduce the spatial smoothening of the polariton wave packet but still it misses the reduction of the total number of particles. Anyway this latter effect is here negligible.
\bibitem{numerical2} We have solved Eq. (\ref{eq:effectiveequation}) by means of a Runge-Kutta algorithm in the time domain exploiting a spatial grid shaped on the group velocity of each layer: in this way, the numerical solution of the propagation is much faster than the one of the three MB coupled equations. In particular, it is able to explore regimes where $v_{gr}/c \ll 1$.
\end{thebibliography}
\end{document}